\documentclass[12pt,a4paper,reqno]{article}

\usepackage{amsmath,latexsym,amssymb,amsthm}
\usepackage{graphicx}
\usepackage{natbib}
\usepackage{amsmath}
\usepackage{amssymb}
\usepackage{amsthm}

\newcommand{\be}{\begin{equation}}
\newcommand{\ee}{\end{equation}}

\theoremstyle{definition}

\theoremstyle{remark}

\numberwithin{equation}{section}

\begin{document}

\title{\bf{Einstein and Tagore, Newton and Blake, Everett and Bohr\\[2ex]\Large The dual nature of reality}}

\author{Anthony Sudbery\\[10pt] \small Department of Mathematics,
University of York, \\[-2pt] \small Heslington, York, England YO10 5DD\\
\small as2@york.ac.uk}

\date{}

\maketitle

\begin{figure}[h]
\scalebox{0.21}{\includegraphics{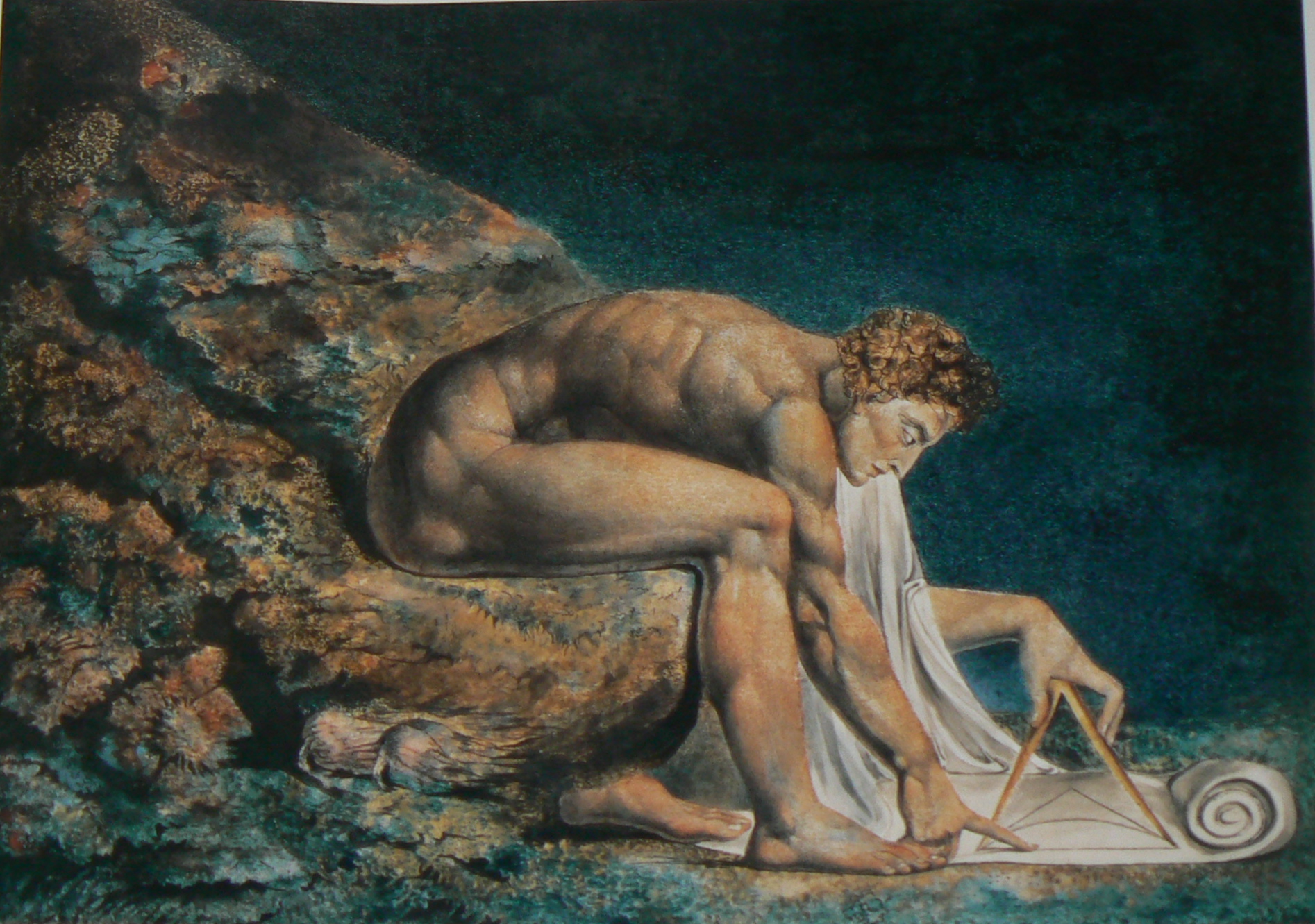}}
\caption{\emph{Newton} by William Blake}
\end{figure}

\newpage

A contribution to the seminar 
\begin{center}
The Nature of Reality: The Perennial Debate\\[5pt]
\end{center}
held at the Indian Institute for Advanced Study, Shimla in March 2012

\vspace{2cm}

\begin{abstract}
 There are two broad opposing classes of attitudes to reality (realist vs idealist, material vs mental) with corresponding attitudes to knowledge (objective vs subjective, scientific vs romantic). I argue that these attitudes can be compatible, and that quantum theory requires us to adopt both of them.

\end{abstract}

\newpage

In their conversation on the nature of reality on 14th July, 1930 (\cite{Einstein-Tagore:conversation}), Albert Einstein and Rabindranath Tagore personified the division between two broad classes of attitude toward the world: one matter-based, centred on an external world in which human beings form a small part, and emphasising the thinginess of things; the other mind-based, centred on human experience and emphasising human creativity and imagination. This characterisation may suggest a division between the scientific and artistic sensibilities, and those are indeed the roles played by Einstein and Tagore in their dialogue; but proponents of the two world views do not separate so neatly according to their pursuits. Moreover, just as scientific and artistic interests are not necessarily antagonistic, so these two attitudes to reality are not necessarily contradictory. I will argue that the effort to achieve a fully scientific understanding of the world impels us to adopt both stances simultaneously.

For Tagore, everything relates to the human mind: ``the Truth of the Universe is human Truth". Einstein, on the other hand, believes in ``the world as a reality independent of the human factor". These seem appropriate positions for the poet and musician on the one hand, working always with immediately felt experience, and the scientist on the other, seeking to detach himself from his fallible senses and attain impersonal certainty by the light of cold reason. Interestingly, though, it is Tagore who attempts to support his position by rational argument, while Einstein simply states his position as a matter of faith: ``I cannot prove that my conception is right, but that is my religion". His part in the conversation concludes ``Then I am more religious than you are!"     

Tagore argues that 
\begin{quote} 
This world is a human world --- the scientific view of it is also that of the scientific man. Therefore the world apart from us does not exist. 
\end{quote}
 This is reminiscent of Derrida's pronouncement ``There is nothing outside the text", which inspired modern (or perhaps post-modern, i.e.\ pre-21st century) sociologists of scientific knowledge like Andrew Pickering, for whom all scientific knowledge is a ``social construction" and who see it as their task to explain what scientists believe without referring to the content of those beliefs. We do not believe in quarks because there really are quarks which scatter electrons in certain ways, but because ... well, I'm not quite sure why, but it's all very sociological (\cite{Pickering}). Pickering believes that the scientist's explanation of our belief in quarks is circular, because he can see no difference between ``Quarks exist" and ``We believe that quarks exist". Many scientists will find this as uncongenial as Einstein did, but well-brought-up quantum physicists may have qualms. Didn't Bohr teach us that there is no world beyond the scale of laboratory apparatus that we have constructed? Doesn't this support Tagore's view, at least in the form ``the world apart from our apparatus does not exist"?

Einstein felt no need to give Tagore arguments in support of his belief in an external reality. He presents it as a matter of faith. It could perhaps be seen as a statement of intent: ``I am determined to understand the world as an external reality".\footnote{For an account of Einstein's realism along these lines, see chapter 6 of \cite{Fine:shaky}.} In these terms he might have seen Bohr's quantum mechanics as an admission of defeat. Margenau commented 
\begin{quote} 
Like most scientists, Einstein leaves unanswered the basic metaphysical problem underlying all science, the meaning of externality [\cite{Margenau:Einstein}],
\end{quote}
but if belief in external reality is understood as a statement of intent, then the belief needs no justification and ``externality" needs no definition; Einstein can simply say ``I will know it when I see it" and go on looking for such an understanding.

If one cannot completely analyse the concept of ``external reality," one can certainly point to particular theories which exemplify it. The paradigmatic example is Newton's dream of a purely mechanical theory: 
\begin{quote}
I wish we could derive the rest of the phenomena of Nature by the same kind of reasoning from mechanical principles, for I am induced by many reasons to suspect that they may all depend upon certain forces by which the particles of bodies, by some causes hitherto unknown, are either mutually impelled towards one another, and cohere in regular figures, or are repelled and recede from one another. [\cite{Newton:dream}]
\end{quote}
 This has inspired the view that physical reality consists entirely of point particles, each of which has a definite position at every instant of time, together with various other numerical properties such as mass and electric charge; the motion of these particles is completely determined by the forces between them. The reality of these particles owes nothing to human minds; it seems to me to be a prime example of what could be meant by a concept of an objective, external reality.

A statement that the physical aspects of reality could be exhaustively described in such a way is not necessarily a denial that there are also human aspects to reality, such as creativity, free will, morality and so on. But it can certainly look like such a denial, and therefore arouses hostility in a romantic, artistic sensibility. This hostility is seen above all in William Blake. Blake exalted freedom and creativity, and regarded reason as something to be fought: 
\begin{quote}
\hspace{3cm} May God us keep\\
From single vision and Newtons sleep. [\cite{Blake:Newton}]
\end{quote}
In his mythology the restrictive, tyrannical God (and creator of the physical world) is called Urizen, which can be heard as ``your reason". 

Blake's picture of Newton (Fig.\ 1) shows this single vision. Newton's objective reality is represented by a mathematical diagram, pale and \emph{un}real compared to the rich, colourful but subjective reality which he is prevented from seeing by his concentration on the objective but abstract, mathematical aspects of reality. Blake does not deny the validity of Newton's vision; it is notable that Newton's mathematical instrument is the same as that held by God (Urizen) in Blakes' picture of the creation of the world (Fig. 2). But he is hostile to the \emph{singleness} of this vision. In the words that Einstein used about quantum theory, he denies its completeness.

\begin{figure}
\begin{center}
\scalebox{0.21}{\includegraphics{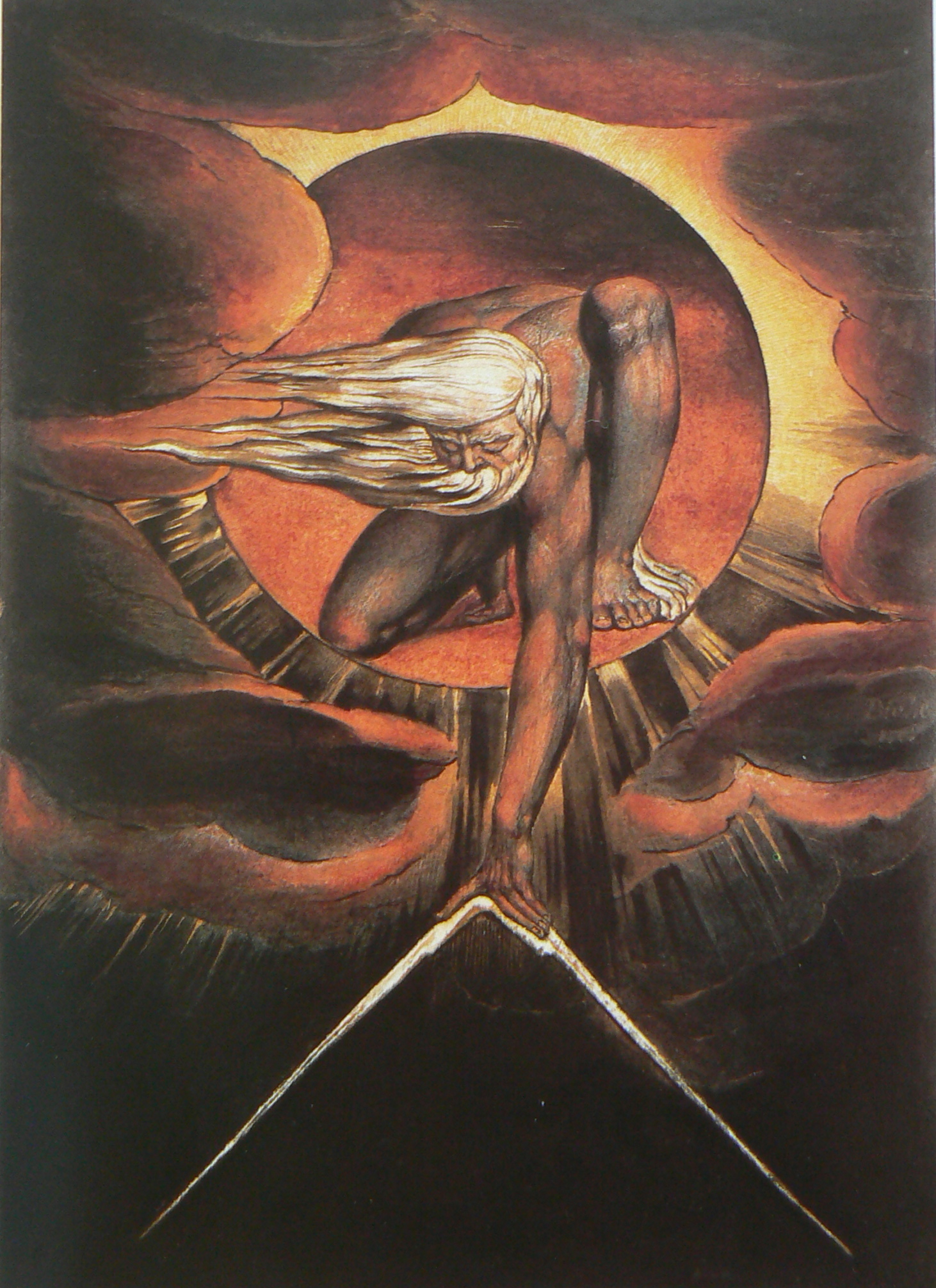}}
\end{center}
\caption{\emph{The Ancient of Days} by William Blake. The creator is Urizen, representing law and reason; an entirely negative figure in Blake's mythology.}
\end{figure}

Einstein's idea of reality had different contents from Newton's, but I think he shared with Newton the idea of reality itself.  They differed as to what kinds of thing are real, but they agreed that there was something real, whether or not it is observed or described, and independent of any observation or description.  Einstein's commitment to realism, in this sense, was a major reason, though not the only one, for his resistance to the claims of quantum theory to be a possible final theory of the world. It is explicitly adduced as the ground for his famous challenge, in the EPR paper (\cite{EPR}), to the completeness of ``the quantum-mechanical description of reality". This argument also requires a principle of locality -- as he later stated it, ``the real states of spatially separated objects are independent of each other" [\cite{Einstein:reply} p.682]. But this is introduced in an almost casual, incidental way at the end of the EPR paper; it is the concept of reality that bears all the weight of the argument. And this was only to be expected; as it had taken shape as a coherent conceptual structure in the hands of Bohr and Heisenberg, the understanding of quantum theory had come to be seen as inseparably linked with an anti-realist attitude to scientific knowledge. Bohr repeatedly insisted that the equations and mathematical objects of the theory were not to be seen as pictures of a microscopic physical reality, but as algorithms for calculating macroscopic effects which are all that humans can presume to know -- a position that was rejected by Einstein, Podolsky and Rosen (EPR) from the outset, in the very words of their title. In 1935, EPR and Bohr both wrote papers entitled ``Can the quantum-mechanical description of reality be considered complete?"  EPR's answer was ``No". Bohr's answer, essentially, was ``What reality?"

It is now generally agreed, following the brilliantly clear light that John Bell shone on the problem, that realism and locality together are indeed incompatible with quantum theory, and that the incompatibility is not a metaphysical matter of opposing concepts, but a clear-cut question of conflicting predictions for the results of experiment. It has been possible to put the question to the test, and the experimental verdict seems (almost, but not quite, inarguably) a clear vindication of quantum theory. 

Unfortunately, Einstein did not live to see Bell's theorem. He would no doubt have been as surprised as everyone else to see that the possibility of completing quantum mechanics could be tested so decisively without any need to consider what form the more complete theory might take. No doubt, also, he would have felt the experimental results as a jarring blow to his world-view. But, perhaps surprisingly, he would not have had to abandon his fundamental faith in the existence of external physical reality. There are in fact two ways in which quantum theory can still be seen as a realist description of the world. Both would have been found uncongenial by Einstein, but perhaps not as uncongenial as abandoning his faith in external reality.

The first realist conception of quantum theory must have been known to Einstein, as it was proposed by Louis de Broglie at the 1927 Solvay conference at which Einstein was present. In de Broglie's theory realism is achieved in just the way assumed in the EPR paper: the quantum description of a collection of particles is acknowledged to be incomplete, and is completed by further elements of reality in the form of precise positions for each of the particles. However, it explicitly violates the principle of locality which was also assumed by EPR. At the Solvay conference it was subject to vigorous attack, in particular by Pauli, in the face of which de Broglie withdrew the theory, though it was later revived by David Bohm who showed that at least some of Pauli's criticisms were unfair. The nonlocality is undeniable, however, and makes it difficult to reconcile this theory with relativity. Even now that nonlocality is accepted by many physicists as an actual feature of the world, there is little support for the de Broglie-Bohm theory.

The second option for realists only emerged after Einstein's death, in 1957, when Everett and Wheeler published their ``relative-state interpretation" of quantum theory (\cite{Everett},\cite{Wheeler}). Here we have an opposite view to EPR; instead of regarding it as incomplete and in need of further elements, Everett and Wheeler take the quantum-mechanical description of reality completely seriously and, indeed, remove from it an element which had always seemed artificial and awkward. This element, known as the ``collapse of the wave function", was probably motivated by realist hankerings and was not found necessary by those who adhered rigorously to Bohr's anti-realist doctrine; but it sits uneasily with any attempt to take quantum theory as a realistic account of the world. However, if the collapse of the wave function is removed from this account, the world --- or rather, according to later accounts, many worlds --- that emerges is so strange that at first few people could accept it. In order to discuss this theory and defend my characterisation of it as a realist view, I must step back and describe the orthodox quantum theory from which it arose.

The following features of quantum theory had become established as common ground by 1935. Associated with any physical system is a mathematical object (its ``wave function" or ``state vector") which changes in time according to a well-defined law (the Schr\"odinger equation) determined by the forces acting on the system. The natural tendency is to regard this as analogous to Newtonian mechanics, in which the physical system would be a collection of particles, the mathematical object is a collection of geometrical points and velocity vectors, and the change in time of this object, determined by the forces acting on the particles, is given by Newton's laws of motion. However, in Newtonian mechanics one can find out the exact mathematical description of the system by observing the positions and velocities of the particles; conversely, a specification of the mathematical description tells one exactly what one would find on observing the system. In the microscopic world of quantum theory, by contrast, it is not possible to observe enough properties of the system to determine its state vector (one sometimes knows the state vector if one has prepared the system in a particular way, but it is not possible to find out how someone else has prepared the system just by observing it); and the mathematical description does not, in general, tell you what you can expect to see on observing the system. It only tells you the possible results of an observation and the \emph{probabilities} of these different results.

This state vector is not usually easy to see as a picture of the system, in the way that the Newtonian mathematical object is a picture of a set of particles in definite positions, moving in definite directions with definite velocities. There is indeed a form of the state vector which corresponds to the particles being in particular positions (though not velocities); but there is another form corresponding to a different configuration of the particles, and there is a third form (the ``sum" of the other two) which somehow contains both configurations. This sum tells us probabilities of finding either of the two configurations when we observe the system, and it is tempting to think of it as containing the information that the system is in \emph{either} the one configuration \emph{or} the other. But this will not do; its logical consequences are definitely wrong (it wipes out the ``interference effects" which show the wavelike nature of matter -- one of the strange empirical findings which showed the necessity for quantum theory in the first place). As a result, it has become common to say that the third (sum) form of the state vector describes a system which is in \emph{both} the first configuration \emph{and} the second. The mathematical apparatus of vectors and vector sums seems to support this: the vector sum of a north-pointing velocity vector and an east-pointing velocity vector is a velocity vector pointing NE, describing motion in which one is \emph{both} travelling north \emph{and} travelling east. But it is certainly hard to picture an object which is both here and there, and it is clear why Bohr held that this mathematical structure should not be seen as any kind of description of a real object. ``There is no quantum world", he said. ``There is only an abstract physical description. It is wrong to think that the task of physics is to find out how nature \emph{is}. Physics concerns what we can \emph{say} about nature." [\cite{Bohr:quote}]

The situation is even worse than this: if the system in question is not isolated, but is part of a larger system (as every actual system is part of the universe), then, in general, the system has no definite state vector. The larger system may be described by a state vector (though not, in general, if it is itself part of a still larger system), but this cannot usually be analysed into unique descriptions of its parts separately. Instead, the overall state vector will be the sum of a number of components, each of which describes one of the parts in a definite state and the other part in an associated state. The two parts are said to be ``entangled".   

I have spoken of ``observing" the properties of a system, which is appropriate language for a realist; there is the physical system, out there, with its properties, and the physicist, separate and detached, quietly observing. Every student of physics is taught that this is wrong; the properties can only be discovered by \emph{measuring} the system in an active experimental intervention, which, on the scale on which quantum theory is relevant, will inevitably alter the very properties one wants to observe. For this reason my account of quantum theory would be frowned at by most physics teachers; wherever I have used the word ``observation", it should be replaced by ``measurement". This word, indeed, has become an important and fundamental part of the theory, with the character of a primitive undefined term in the fundamental postulates. In its most austere (and anti-realist) form, the theory renounces any pretension to describe a physical world; it only offers a method for calculating the results of laboratory procedures, each consisting of a specified preparation followed, after a specified lapse of time, by a specified measurement. The calculation will tell us the possible results of this measurement and the probabilities for each of them. 
  
This is neither satisfying nor satisfactory. It is not satisfying because, to many scientists, it misses the whole point of doing science in the first place. Why would we want to predict the results of experiments? We want to understand the world, and we do experiments to check if we've got it right (see \cite{QMPN} p. 214). As John Bell put it, 
\begin{quote}
To restrict quantum mechanics to be exclusively about piddling laboratory operations is to betray the great enterprise" [\cite{Bell:piddling}]. 
\end{quote}
It is not satisfactory because it is hardly possible to obey the stern injunction to consider only laboratory procedures consisting of a single preparation followed by a single measurement. Often one wants to follow the progress of events beyond the final measurement. In that case one has to treat the measurement as a second preparation, giving a new state vector for the system which will serve for the start of a new calculation. But the nature of this preparation, and therefore the identity of the new state vector, depends on the result of the experiment, which is a random outcome over which the experimenter has no control. If we follow the state vector after the first preparation, through the measurement, and beyond, we see an evolution which changes abruptly and randomly at the time of the measurement. This is incorporated in the presentation of quantum mechanics in most textbooks, which (following the original formulation of Paul Dirac (1930) and John von Neumann (1955)) stipulate that the state vector changes in two distinct ways:
\begin{enumerate}\item Left to itself, the system changes smoothly according to the Schr\"odinger equation; but
\item If the system is subjected to a measurement, its state vector changes instantaneously and unpredictably to reflect the result of the measurement. It undergoes ``projection".
\end{enumerate}

Physicists have always felt uneasy about this. It is supposed to be a fundamental statement of a basic law of nature, on a par with Newton's laws of motion. But it lacks the unity and simplicity of Newton's laws: why are there two quite different laws applying in different situations? And anyway, what are these different situations? What is this ``measurement" which enters into the second law as an undefined term? Surely measurements are not basic constituents of the world. An actual measurement in a laboratory is conducted with apparatus which can be analysed as a physical system like any other, and must itself follow the Schr\"odinger equation as in the first postulate. 

This is the measurement problem of quantum mechanics. Von Neumann investigated it by taking the measurement seriously as a physical process, and considering the quantum description of a total system consisting of the object being measured together with the measuring apparatus, which interact by means of known forces between them. Following the progress of the measurement according to the Schr\"odinger equation yields a total state in which the object and the apparatus are entangled: the total state vector is a sum of states in each of which the apparatus shows a definite result and the object definitely has the property shown by that result. But this total state vector does not correspond to any particular result. In order to reflect the fact of experience that a measurement does have a definite unique result, von Neumann found it necessary to appeal to the second (projection) postulate after all.

At this level it becomes very tempting to interpret the quantum sum of state vectors in terms of ``or" rather than ``and". In the laboratory it is surely true, after a measurement has been made, that it had one result \emph{or} another. I said earlier that this interpretation leads to conflict with experimental facts, in that it predicts that there will be none of the interference effects which are characteristic of quantum phenomena (and which are predicted by the Schr\"odinger equation). But for large things like laboratory apparatus, these interference effects are far too small and infrequent to be ever observed; it will never lead to conflict with experiment to assume that the quantum $A + B$ means that either $A$ happened or $B$ happened. This result is known as ``decoherence"; it is well established in theory (\cite{Schlosshauer}).  Nevertheless, it remains an approximate statement; in principle, if it is taken as an exact statement about the apparatus and the object that its state vector is the sum of different results, it cannot be understood as stating that one of these results happened.

In 1957 Hugh Everett challenged von Neumann's conclusion. Why can't we believe, he asked, that the world is described by a state vector which is the sum of components describing situations which we would recognize as different and incompatible states of affairs? The immediate answer is that the world just isn't like that --- we see that it isn't; we never see such sums. In terms of Schr\"odinger's famous example (\cite{Schrcat}), we never see a cat in a state which is the sum of being alive and being dead. Everett's reply (anticipated by Schr\"odinger himself) was that the theory \emph{tells} us that we will never see such a cat. Since we are ourselves physical objects, the universal state vector must describe our brains as well as everything else. If I look at Schr\"odinger's  cat after it has been in his diabolical box for a while, the physical process by which I see it leads, via Schr\"odinger's equation, to an entangled state which is the sum of a state in which the cat is dead and I see it as dead, and a state in which it is alive and I see it as alive. Nowhere is there a state of my brain seeing a cat which is the sum of alive and dead. Everett made the analogy with Copernicus's revolutionary statement that the earth moves round the sum and rotates on its axis. To the objection that it doesn't \emph{feel} as if we're whirling around in space, Copernicus could reply (though he may have had to wait for help from Galileo and Newton) that the laws of physics show that what it feels like to be whirling round in space (on a massive gravitating planet) is exactly what we do feel.

Following Everett, there has been a growing body of opinion among physicists in favour of taking quantum theory seriously and literally: it gives us a mathematical object (the state vector) which constitutes a complete description of the physical world. This state vector has many parts, each of which we can recognize as a picture of a world that we can understand --- it contains atoms and molecules in combinations which we know as planets, and mountains, and trees and tigers and people. But different things are happening in each of these parts: in some of them Schr\"odinger's cat is alive, in some it is dead, in some of them nobody has heard of Schr\"odinger's cat because the young Schr\"odinger decided to become a poet rather than a physicist. The theory contains ``many worlds".

If this is right, our most successful physical theory gives us an account of reality which contains far more than what we see, or could ever see. If I do an experiment which has two possible outcomes, $A$ or $B$, depending on whether a certain radioactive nucleus has decayed or not in the course of the experiment, then what I see is that the experiment went one way (say $A$) and not the other ($B$). I believe that $A$ happened and not $B$ (because I saw it); everyone I ask to check my experiment agrees that I am right; surely, by all the standards of scientific truth, my belief is justified and I am entitled to say that in the \emph{real} world, $A$ happened and $B$ didn't. Yet the theory tells me that there is another equally real world in which $A$ didn't happen but $B$ did.

So why don't we look again at the meaning of the quantum description, with its many worlds? We can have a mathematical description containing all these worlds, and believe that it is really true, without believing that each of the worlds is real. Maybe the vector sum occurring in the mathematics should be interpreted as a disjunction: the statement is that all of these worlds are possible, but only one is real. The problem with this is that it requires a precise definition of what will count as a world, as a matter of basic principle, and this goes against the whole spirit of the theory. We can look at a particular state vector of the universe and say ``Oh yes, this seems to have a lump here which looks like the kind of world we know, with people doing experiments and getting unique results; and there's another lump there with people getting different results; and over there is a smudge which doesn't look like anything much; but they're all combined together with this vector sum idea"; but this interpretation will always be \emph{ad hoc}, and we can't give a general rule for recognising worlds which will apply in advance to any state vector. Those who think this is a problem call it the ``preferred basis problem" of Everett's theory.

There seem to be two conflicting accounts of reality here, both authorized by science. On the one hand, science is based on empirical evidence; what is revealed by careful experiment is real. On the other hand, it advances by taking its successful theories seriously; the theoretical entities of a fundamental theory are real. This tendency has often been resisted by the cautious --- many physicists at the end of the nineteenth century refused to believe in the reality of atoms -- but it seems to me that the general lesson of history is that reality favours the bold.

It is not, of course, the first time that a general scientific theory has seemed to be in conflict with intuition. I have already mentioned the conflict between the theory that the earth spins on its axis and the obvious fact that the sun goes round the earth once a day. Rutherford's discovery that atomic nuclei are so small that atoms are mainly empty space seems to falsify our perception that the things around us are solid. The lesson of relativity that physics takes place in space-time is said to show that there is no such thing as the passage of time. Both determinism and indeterminism are thought, not necessarily by different people, to make free will impossible.

Scientists often respond to these conflicts by saying that science has revealed that the intuitive belief is an ``illusion". We have the illusion that the sun goes round the earth; the solidity of stones, the passage of time, free will -- they are all illusions. In nearly every case I think this is a mistake. It is not an illusion that the sun goes round the earth; Einstein taught us that we are free to adopt a frame of reference in which the earth is fixed (intuitively, we can hardly help doing so), and in that frame the sun does indeed go round the earth. It is not an illusion that stones are solid, it is part of the meaning of solidity; and it is certainly not an illusion that we have free will, it is a clear fact of everyday life (once again it is a matter of the meaning of words, though not an easy one to tease out in this case). In both these cases the problem arises from confusing a clear everyday concept with an unjustified pseudo-scientific theory of that concept: solidity does \emph{not} mean that matter occupies a mathematical continuum, free will does \emph{not} mean that we have the power to interrupt the laws of physics. As for the idea that time is an illusion, which was even held by Einstein, I have no idea what that is supposed to mean. (What is that we mistakenly believe when we are under this illusion?)

In all these cases I believe the resolution of the conflict is not that one of the conflicting ideas is mistaken or illusory, but that there is no conflict. I would like to apply the same strategy to the conflict between the many worlds of quantum theory and the one world of experience; but it will involve some re-examination of our conception of reality.

I see this conflict as one of a wide class of philosophical problems which was identified by Thomas Nagel (1986). These problems arise whenever we attempt to go beyond our own individual situation and experience to obtain a more general, objective account of the world -- something that we may feel obliged to do for ethical, political or aesthetic  as much as for scientific reasons. We want to lay aside our individual interests and work for the general good; go beyond our subjective opinions and attain objective knowledge; move out of our particular situation and see the overall picture, or, as Nagel puts it, attain a ``view from nowhere" as opposed to the ``view from now here". The problem is that the objective propositions we reach, which we seem to have good reason to believe, then appear to conflict with the subjective experience which is still vividly present to us. Nagel acknowledges the force of the objective position, but rejects the temptation to dismiss the subjective experience as merely an illusion; the subjective experience has a vividness and a reality compared with which the objective truth is abstract, pale, ``etiolated".

The contrast is beautifully depicted in Blake's picture of Newton (Fig.\ 1). Newton is depicted as a thoroughly physical, muscular young man, but he is totally concentrated on the abstract mathematical laws of nature. He sees only the thin, colourless diagram on his scroll, and does not see the colour, texture and concrete reality of the coral-encrusted rock on which he sits. He sees all of reality; as we have already noted, the dividers he holds are the same instrument as that held by God in Blake's picture of the creation of the world (Fig.\ 2). But he also misses all of reality, in the richness behind him.\footnote{As I was writing this talk, the radio was playing music chosen by an explorer. The music stopped, and I heard him explaining the purpose of his expeditions into the rain forest. ``I was not there as a scientist", he said, ``I was there to \emph{understand} the place." This paradox --- aren't ``science" and ``understanding" synonymous? --- nicely captures the two sides of Blake's picture.}  

Now, in the farthest we have travelled along the journey initiated by Newton, we find the reality of subjective experience being forced on us by the abstract theory itself. If we start as fundamentalist physicists, we have a quantum-mechanical state vector which we think describes the whole of physical reality at different times. Suppose we find that at a certain time this state vector contains a world in which there is a scientist preparing Schr\"odinger's experiment, putting a cat in a box with a radioactive nucleus and a phial of volatile poison. At a later time we will see that this part of the state vector has split into two: one part containing a dead cat and the scientist with his brain registering the thought ``Oh dear, the cat died", and another part containing a living cat and the scientist's brain registering the thought ``Ah, the cat is still alive". From our external viewpoint, we see that neither of these thoughts is a true reflection of reality. But the thoughts do not occur in our external framework; each of them belongs in its own part of reality, and in that context it fulfils all the conditions for truth: it corresponds to an actual physical fact, it agrees with the thoughts of all other competent observers in that part of reality, and so on. Although the two thoughts are contradictory, each of them is true in its own branch of reality. It is like the truth of a sentence in a work of fiction (\cite{Lamarque}); the sentence takes its truth from the story in which it belongs. In St. Luke's gospel, it is true that Christ died on the cross; but in Kazantzakis's novel \emph{The Last Temptation of Christ} it is true that he comes down from the cross and marries Mary Magdalene. A similar approach to truth and reality can be used to counter the sceptical thought ``Maybe none of this is real; maybe it is all a dream." If it is a dream, then the thought ``This table is real" is occurring in the dream, and as such is true: the table is real.

So there is one nature of reality for the many worlds described by quantum theory, another (no less real) for the one world we actually experience. I like to think that the first view of reality reconciles quantum theory with Einstein's belief in a real external world independent of human observers (though I cannot be confident that Einstein would have agreed). The second (one-world) type of reality, however, explicitly depends on a particular observer for its definition. I believe that the phenomenon of decoherence makes it possible to replace this single observer with a community of observers, but still, the scientific definition of this genus of reality is linked to consciousness. Not necessarily human consciousness, in principle; nevertheless, this interpretation of quantum mechanics is remarkably close to Tagore's view. Tagore could have been referrring to Everett's paper when he said\footnote{in a part of the conversation with Einstein which is not contained in \cite{Einstein-Tagore:conversation}, but was unearthed by Partha Ghose} 
\begin{quote}
[the world] is a relative world, depending for its reality upon our consciousness.
\end{quote}

This is why my title has (three times) the word ``and" (rather than ``versus"). To understand quantum theory, we seem to need Einstein's reality \emph{and} Tagore's, Newton's \emph{and} Blake's, Everett's \emph{and} Bohr's. 

What are we, living in one definite world, to make of the quantum description of many worlds in a vector sum? How does it affect us? At the start of his experiment with the cat, Schr\"odinger knows the quantum state vector that describes how he has set up the experiment, and he can calculate what it will become at the end of the experiment. It will then describe a world in which the cat is alive, and another in which it is dead. What does this calculation mean for Schr\"odinger (\emph{a}) at the start of the experiment, when it lies in the future, (\emph{b}) after the experiment, when it lies in the present?

(\emph{a}) Everybody agrees on how to use a quantum calculation. Before the experiment, the different worlds in the future state vector describe the different possible results. More than that, the relative sizes of these components give the probabilities for these different results.

(\emph{b}) Let's look on the bright side and suppose that the cat lives. Schr\"odinger has the immediate reality of the living cat in front of him; he also knows that there is another world with a dead cat in the quantum description. What can that mean, in his lived reality? He must take it as describing something that might have happened, but didn't. However, he can't cheerfully say ``Oh good, it didn't happen; let's stop worrying about it" and start again with a state vector describing the living cat; that would be to ignore his own equation and apply the projection postulate, which gives wrong answers. The chance of an error is tiny, but in principle it is there. To calculate the results of future experiments, he must take into account the fact that the cat might have died.  What might have happened, but didn't, can still affect the real world; therefore it is still a part of reality.  

Einstein had two other independent objections to quantum theory. One of them was his belief in determinism. Realism and determinism are logically independent --- a theory could exhibit either of them without the other --- but it is not always clear that these were distinct for Einstein. In 1949 he was protesting 
\begin{quote}
I still believe in the possibility of a model of reality -- that is to say, of a theory which represents things themselves and not merely the probability of their occurrence. [\cite{Einstein:reply}, p. 669].
\end{quote}
In analysing this statement, one might note that it is not things that occur but events, and in asking for events themselves to be represented in the theory, without probabilities, Einstein seems to want all of space-time (past, present and future) to have definite and univocal existence in the theory. Such a theory need not be deterministic, in the sense that identical causes must always be followed by identical effects, but it does require that the future is fixed (one might say it is ``fatalistic"). In any case, in requiring no probabilities, Einstein does seem to be asking for determinism in the strong sense. As is well known, he stated elsewhere that ``I, at any rate, am convinced that He [the ``old one"] does not throw dice" [\cite{Einstein:dice}].  

In this respect also, Everettian quantum theory can be seen as meeting Einstein's desiderata (though, again, I'm not sure that Einstein would have been happy with it). The objective, realistic aspect of the theory --- the ``view from nowhere" --- is fully deterministic, as the state vector of the universe obeys the Schr\"odinger equation. Given the state vector at any one time, the state vector at any future time (and, indeed, any past time) is completely determined. But, of course, the state vector relative to any given consciousness is subject to random changes; this aspect of the theory is completely indeterministic. The ``old one" does not play dice with the whole world, but he does play dice with each one of us. In the view from ``now here", the future is open (\cite{openfuture}); the state vector of the universe gives us only a set of possibilities for our future, together with probabilities for each of them, but at any given time there is no such thing for us as \emph{the} future. This makes it necessary to rethink what we mean, in this internal perspective, by probability and the truth of statements in the future tense (\cite{phillessons}).

In the internal perspective, present truth and reality are guaranteed by my perceptions (``I" being the observer whose consciousness defines the perspective). Clearly, there can be no such guarantee from experience for the future. Nevertheless, it seems natural to believe that there is such a thing as my future experience, even though I don't know what it is; that there is a true statement about what my future state will be. This is denied by the quantum calculation: there is nothing to link any one of the components of the future state vector with the particular component that I am experiencing now. It might also be denied by a naive view, perhaps our own view as children (``of course there isn't a definite future, it hasn't happened yet; anything could happen") and it was certainly denied by Aristotle, who, in his famous passage about the sea-battle (\cite{seabattle}), asserted that statements about the future are neither true nor false. This has been taken by many logicians as meaning that bivalent logic does not apply to future-tense statements; there is a third truth value as well as ``true" and ``false". But Aristotle also pointed out that statements about the future can be more or less likely. This suggests that their truth value lies on a scale between 0 and 1, and should be identified with probability.

So from the internal perspective, there is one component of the present universal state vector which represents the truth now, and present-tense statements have truth values 0 or 1 determined by their consonance with this component. Future-tense statements have truth values equal to the probabilities calculated by quantum mechanics. 

This suggests further elaboration of the external perspective. In this perspective the universal state vector represents the whole truth about reality. If this is the sum of many components describing worlds that we could recognize, then it is often said that these many worlds are ``all equally real". But there is no warrant for that ``equally". These components do not have equal lengths as vectors; if that length should happen to be zero, then the corresponding world is not real. A component with a length which is tiny but not actually zero, however, is supposed on this view to be fully real. It seems much better to say that full reality belongs only to the actual universal state vector; any other vector (for example, one of the components representing a recognisable ``world") has a lesser degree of reality measured by its contribution to the universal state vector. In symbols, if the universal state vector is 
\[
\Psi = c_1\Psi_1 + c_2\Psi_2 + \cdots
\]
where $\Psi_1,\Psi_2,\ldots$ describe different recognisable worlds, then the degree of reality of the world $\Psi_1$ is $|c_1|^2$, that of $\Psi_2$ is $|c_2|^2$, and so on.   

In the external perspective, Everettian quantum mechanics is a realistic, deterministic theory --- just what Einstein wanted. I feel pretty sure, however, that he would not have been satisfied with it. The problem is that neither the external perspective nor the internal one have a third feature that Einstein regarded as essential in a scientific theory, namely locality, or more generally separability:
\begin{quote} Now it appears to me that one may speak of the real factual situation of the partial system $S_2$. \ldots But on one supposition we should, in my opinion, absolutely hold fast; the real factual situation of the system $S_2$ is independent of what is done with the system $S_1$, which is spatially separated from the former. [\cite{Einstein:autobiog} p. 85]
\end{quote}
The existence of entanglement means that it is not possible to divide the world into pieces in such a way that the real factual situation of the whole world is completely described by putting together the real factual situations of its parts; indeed, the parts do not have their own real factual situations. In particular, one cannot divide space-time into regions and describe these regions independently of each other. The theory is \emph{nonlocal} or \emph{holistic}.\footnote{I should add that this is currently a matter of controversy. I have been following Everett in using the Schr\"odinger picture as a description of reality. Deutsch and Hayden (\cite{Deutsch-Hayden, Deutsch:reply}) claim that the Heisenberg picture gives a fully separable description of the world, but it is not clear that this is valid (\cite{Wallace-Timpson}).}

\section*{Conclusion}

It has often been remarked that there is little engagement between Einstein and Tagore in their conversation on the nature of reality. They both state their views, which just sail past without affecting each other. The sole discussion between Everett and Bohr was described by Everett's biographer as ``simply a polite hearing and a lot of mumbling" and by Everett himself as ``that was a hell of a---doomed from the beginning" (\cite{Byrne}, pages 168 and 221). It is hard to even imagine a meeting between Newton and Blake. My thesis here has been that although these incompatibilities are real, we do not have to choose one side or the other: we can understand the different contexts in which notions of truth and reality apply, and we can place ourselves in either context at will, without inconsistency. Quantum theory forces us to consider both; and each context shows us a different nature of reality.


\bibliography{quantum}
\bibliographystyle{authordate1}

\end{document}